 \documentclass[nohyper,12pt,letterpaper]{JHEP3}
 \usepackage{graphicx}

 \def \ha{\frac{1}{2}}
 
 \title{The Regge Limit for Green Functions in Conformal Field Theory}
 \author{T.\,Banks\\
 Department of Physics and SCIPP\\
 University of California, Santa Cruz, CA 95064\\
 E-mail: \email{banks@scipp.ucsc.edu}\\
 {\it and}\\
 Department of Physics and NHETC, Rutgers University\\
 Piscataway, NJ 08540}

 \author{G.\,Festuccia\\
 Department of Physics and SCIPP\\
 University of California, Santa Cruz, CA 95064\\
 E-mail: \email{guido@scipp.ucsc.edu}\\
 }

 \abstract{We define a Regge limit for off-shell Green functions in quantum field theory,
 and study it in the particular case of conformal field theories (CFT). Our limit differs
 from that defined in \cite{port}, the latter being only a particular corner of the Regge regime.
 By studying the limit for free CFTs, we are able to reproduce the Low-Nussinov\cite{ln},
 BFKL\cite{BFKL} approach to the pomeron at weak coupling. The dominance of
 Feynman graphs where only two high momentum lines are exchanged in
 the t-channel, follows simply from the free field analysis. We can
 then define the BFKL kernel in terms of the two point function of a
 simple light-like bilocal operator.  We also include a brief discussion of the
 gravity dual predictions for the Regge limit at strong coupling.}

 \received{} \accepted{}
 \preprint{hep-th{}\\SCIPP-2009/15\\RUNHETC-2009-24 \\}
 
\begin{document}

\section{Definition of the Regge limit}

In the grand old days of Regge theory, a great deal of effort was devoted to finding Regge behavior of amplitudes
in field theory by summing classes of Feynman diagrams\cite{sum}.  There was, to our knowledge, no satisfactory
{\it derivation} of Regge pole behavior by these methods, because it was not possible to show that the neglected
diagrams were indeed sub-leading in the Regge limit.

The modern QCD based approach to the Regge limit \cite{BFKL} studies the limit $s \rightarrow \infty$ with $-t \gg
\Lambda_{QCD}^2 $ fixed. It is claimed that leading order weak coupling results are valid in this regime. In the
case of the scattering of quarkonia, where both target and projectile are composed of quarks with mass $ \gg
\Lambda_{QCD}$ this seems to be correct\cite{mueller}. Nonetheless, there does not seem to be a clear
understanding of the corrections to leading order formulae, the domain of applicability of the results, or their
relevance to actual experiments. Furthermore, these results explore only a corner of the full Regge regime, which
is not the relevant regime for most phenomenology.

A new slant on this subject was provided by the seminal work of \cite{bpst}, which used the AdS/CFT
correspondence. These authors observe that at least in the planar limit at strong 't Hooft coupling, the string
dual of ${\cal N} = 4$ Yang-Mills theory, defines a Regge limit, in which one can calculate world sheet integrated
correlation functions of string vertex operators. These should correspond to planar
correlators of CFT operators\cite{Maldacena:1997re}, so there should be a Regge limit in conformal field theory!  The authors of
\cite{bpst} did not provide a precise definition of the Regge limit for CFT correlation functions, which was
attempted by \cite{port}.

The aforementioned work motivated us to look for a definition of the Regge limit purely in CFT terms. We find
something a bit different than the authors of \cite{port}, and we will explain the differences below. Our definition of the Regge limit can be applied to a completely general quantum field
theory. In this paper we will explore only conformally invariant examples, where the results are relatively
simple. With luck, this line of research might lead to a rigorous framework for studying Regge behavior in real
QCD.

Consider a generic time ordered Green function

$$ \langle 0 | T \phi_1 (x_1) \phi_2 (x_2 = 0) \ldots \phi_n (x_n ) | 0 \rangle .$$ The fields can be either
elementary or composite at each point. Despite the notation, they need not all be scalars. Note that we have used
translation invariance to bring the second field to the origin. The Regge limit is defined by choosing a boost
leaving the origin fixed, and applying its unitary representative, $U(\omega )$ {\it only to fields at the odd
numbered points}. That is $$G \equiv \phi_{2i + 1} (x_{2i+1} ) \rightarrow U^{\dagger} (\omega)\phi_{2i + 1}
(x_{2i+1} ) U (\omega).$$ $\omega$ is the rapidity of the boost, and the Regge limit is the limit $\omega
\rightarrow\infty$.

If we consider the four point function in a theory with mass gap, our definition does not change the masses of the
external legs, and therefore it can be translated to the momentum space invariant amplitude on mass shell. It
corresponds to boosting $p_1$ and $p_3$ (and applying the appropriate boost matrix to the spin indices of
$\phi_{1,3}$ ) with $p_2$ fixed and $p_4$ determined by momentum conservation\footnote{For a two body elastic
process in the center of mass frame $p_4$ stays on-shell if the boost is in a direction transverse to the
exchanged momentum $p_1-p_3$}. Thus $ s = (p_1 + p_2)^2 \rightarrow e^{\omega} s_0$ and $t = (p_1 - p_3)^3$ is
fixed, and we reproduce the traditional Regge limit.  The Regge limit is thus the limit where some fields in a
Green function are given a large boost relative to others. The boost is chosen to leave the coordinate of one of
the unboosted points invariant. The boost singles out a 2-plane in Minkowski space, which we will henceforth refer
to as {\it the boost plane}.

We note that momentum space amplitudes, with momentum conservation imposed, are automatically translation invariant. Since boosts do not commute with translations, this introduces an ambiguity into the choice of boost that goes into the definition of the Regge limit of coordinate space Green's functions. Despite the fourfold continuous infinity of choices, there are really only two distinct limiting regimes. One can either choose, as we have, a boost which leaves one of the points in the correlation function invariant, or a boost around a generic point.  The authors of \cite{port} define a Regge limit for CFT 4-point functions in terms of cross ratios.   If
the cross ratios are parametrized as $z\bar{z}$ and $(1 -z) ( 1 - \bar{z})$, then the limit takes both $z$ and $\bar{z}$ to zero, with fixed ratio. This is what happens if one does a relative boost of $x_{1,3}$ and $x_{2,4}$ with a boost that leaves a generic point fixed.   We prefer the choice of leaving one of the correlation points fixed because, as we now show, this coincides exactly with the behavior of momentum space Green's functions.   In free field theory, the prescription of \cite{port}, gives a different scaling for coordinate space and momentum space Green's functions, and the relative exponents for the leading behavior depend on which Feynman graph one is computing, and also on the  spins of the underlying free fields and the choice of composite operator.  As we will see below, our definition of the Regge limit for 4-point functions corresponds to taking one cross ratio to zero with the other fixed.

\subsection{Translations and Fourier Transforms}

Consider $m+n$ points $x_i^{\mu}$ for $i=1,...,n+m$. Boosting the last $m$ points gives:
$$
(x_1,...,x_{n-1},x_n,...,x_{n+m-1})\rightarrow (x_1,...,x_{n},\Lambda x_{n+1},...,\Lambda x_{n+m})
$$
Now we can ask what happens by first translating the points by $y^{\mu}$ then boosting as above and finally undoing the translation the result is
$$
(x_1,...,x_{n-1},x_n,...,x_{n+m-1})\rightarrow (x_1,...,x_{n},\Lambda x_{n+1}+z,...,\Lambda x_{n+m}+z)
$$
where $z=(\Lambda-I)y$.
Therefore if we start with a function $f(x_i)$ which is invariant under simultaneous translation of all the arguments by the same amount and apply the boosting procedure we end up with a function which does not have this property anymore.
As a consequence the Fourier transform of the transformed function will not be proportional to the delta function of the sum of the $n+m$ incoming momenta.

Now consider first translating to $x_1=0$ and then boosting the result is:
$$
(x_1,...,x_{n-1},x_n,...,x_{n+m-1})\rightarrow (0,x_2-x_1..,x_{n}-x_1,\Lambda (x_{n+1}-x_1),...,\Lambda (x_{n+m}-x_1))
$$
The transformation is clearly compatible with translation invariance of the original function because the transformed function will depend only on the differences $x_i-x_1$. In order to see more explicitly what happens with Fourier transforms it is better to introduce a $x_1'$ variable in a way compatible with the translation invariance just established:
$$f'(x_1',x_2',..,x_{n+m}')=f(x_1',...,x_{n}',\Lambda^{-1}(x_{n+1}'-x_1')+x_1',...,\Lambda^{-1}(x_{n+m}'-x_1')+x_1')$$
Indeed the right hand side is equal to $$f(0,x_2'-x_1',..,x_{n}'-x_1',\Lambda^{-1}(x_{n+1}'-x_1'),...,\Lambda^{-1}(x_{n+m}'-x_1'))$$
while the left hand side is by definition $f'(0,x_2'-x_1',...,x_{n+m}'-x_1')$ so that:
$$
f'(0,x_2'-x_1',...,x_{n+m}'-x_1')=f(0,x_2'-x_1'..,x_{n}'-x_1',\Lambda^{-1}(x_{n+1}'-x_1'),...,\Lambda^{-1}(x_{n+m}'-x_1'))
$$
which is the action of the transformation we are considering.
Start with the Fourier transform of $f(x_i)$:
\begin{eqnarray}
&&\int \prod_i dx_i e^{\sum_i x_i\cdot p_i}f(x_1,...,x_{m+n})=\cr=&&\int \Pi_i dx_i e^{\sum_i x_i\cdot p_i}f'(x_1,...,x_n,\Lambda(x_{n+1}-x_1)+x_1,...,\Lambda(x_{n+m}-x_1)+x_1)
\nonumber
\end{eqnarray}
Now change variables to $x_{n+i}'=\Lambda(x_{n+1}-x_1)+x_1$ getting
\begin{eqnarray}
\int \prod_{i=1}^n dx_i\prod_{j=n+1}^m dx_{j}' e^{i \left(p_1+\sum_{j=n+1}^{n+m}(I-\Lambda)p_j\right)\cdot x_1} e^{i\sum_{k=2}^n p_k\cdot x_k+i \sum_{k=n+1}^{n+m}\Lambda p_{k}\cdot x'_{k}}f'(x_1,..,x_n,x'_{n+1},..,x'_{n+m}) \nonumber
\end{eqnarray}
This is the fourier transform of the function $f'$ with momenta
\begin{eqnarray}
&&p_1'=p_1+\sum_{j=n+1}^{n+m}(I-\Lambda)p_j=-(p_2+...+p_{n}+\Lambda p_{n+1}+\Lambda p_{n+m})=-\sum_{k=2}^{n+m}p_k' \nonumber\\
&&p_2',...,p_{n}'=p_2,...,p_n \nonumber\\
&&p_{n+1}',...,p_{n+m}'=\Lambda p_{n+1},...,\Lambda p_{n+m} \nonumber
\end{eqnarray}
 Which is the natural definition of the Regge limit in momentum space for off-shell Green functions as discussed above.
For four point functions generically this transformation does not keep $p_1^2$ invariant but for any value of on shell $p_i$ there exist a class of $\Lambda$ which keep the particles on the mass shell. We conclude that our definition of the Regge limit of coordinate space Green's functions, employing a boost which keeps one point in the Green's function fixed, will always coincide with the Regge limit in momentum space, for all operators in all field theories.

\subsection{Light cone quantization and t-channel exchange}

Now let us consider quantization of the field theory in a light cone frame with $x^+ \rightarrow e^{\omega} x^+$,
under the Regge boost. In this frame, all the odd numbered points will be later than all the even numbered ones.
Using Lorentz invariance of the vacuum, we can rewrite the Green function as

$$G = \langle 0 | T_+ [\phi_1 (x_1) \ldots \phi_{2k + 1} (x_{2k+1})] U(\omega ) T_+ [\phi_2 (0) \ldots \phi_{2k+2}
(x_{2k + 2}) ] | 0 \rangle .$$ Here we have assumed $n = 2k+2$.  If, instead, $n = 2k + 1$ then the last field has
$2k + 2 \rightarrow 2k$. Thus, in the light cone frame, the Regge limit defines a {\it channel}, splitting the
operators in the correlator into two groups separated by a boost. For the four point function, this is just
Mandelstam's $t$ channel. This description allows us to write the correlator in the following suggestive manner:

\begin{equation}\label{Up} G = \langle S_o | U(\omega ) | S_e \rangle = \int\ d\lambda\ \rho (\lambda) \psi_o^*
(\lambda ) \psi_e (\lambda ) e^{i\omega\lambda} ,\end{equation}
 where $\rho (\lambda )$ is the spectral density of the boost,
and the wave functions are overlaps with states of fixed {\it boost eigenvalue}.  We can think of the boost as a
Hamiltonian, and the Regge limit as the long time evolution under that Hamiltonian.

From a purely mathematical point of view, if the product of wave functions and spectral density is a smooth
function, $f$, of the real variable $\lambda$, then the Riemann-Lebesgue lemma tells us that the correlator falls
off faster than any power of $\omega$. Indeed, we expect a behavior of the form

$$G \rightarrow e^{ - \alpha \omega} \omega^k ,$$ where $\alpha $ is the imaginary part of the nearest singularity
of $f$ in the complex $\lambda$ plane, and $k$ is determined by the nature of the singularity.  A simple pole
leads to a pure exponential, which is the traditional prediction of Regge theory.  It is well known from the work
of \cite{BFKL} that in weakly coupled Yang-Mills theory, the behavior is more complicated, and has $k \neq 0$,
which generically corresponds in old fashioned language to a Regge cut. The authors of \cite{bpst} have argued
that Regge pole behavior will be restored in a confining theory, but at least in the vacuum channel, this is at
best valid in the planar limit. The Pomeron pole behavior coming from the exchange of vectors is inconsistent with
unitarity, and is generally expected to be replaced by a cut. Data on total cross sections also suggest this.

It is important to note that in the above derivation, we took the plus components of BOTH $x_{1,3}$ positive.   After the boost, this does not correspond to the causal ordering we expect for scattering amplitudes.  $x_1$ is in the future of $x_4$.  Scattering kinematics would be preserved if we took $x_1^+$ negative.  Explicit computations in free field theory show that the scaling behavior of Green's function in the Regge limit is independent of the sign of $x_1^+$.  We will show later that, for a line of weakly coupled fixed points, the Regge limit of four point functions is defined by a Bethe-Salpeter like equation (as in the work of BFKL) of the form
$$ G = G_0 + \int K G ,$$
where the kernel $K$ is exchanged in the t-channel. As a consequence of Lorentz invariance of the integration measure in this equation, the result of a relative boost between $x_{1,3}$ and $x_{2,4}$ is simply a change in the kernel. We believe that a proof that the Regge exponent is independent of the sign of $x_1^+$ could be constructed by analyzing the analytic structure of this kernel.  It certainly seems to be true order by order in perturbation theory.  Thus, our simple expression for the limiting Regge behavior in terms of the matrix element of a boost operator, should also control the Regge behavior of scattering amplitudes. More work is needed to clarify the required analytic continuation in a way that does not depend on perturbation theory.

It is our belief that the old fashioned \lq\lq derivation" of Regge behavior from a Sommerfeld-Watson transform of
a non-convergent partial wave expansion is somewhat misleading.  Pure Regge pole behavior is a mathematical
ansatz, with no intrinsic physical principle underlying it.  Our derivation of Regge behavior is much more robust,
since a wide variety of singularities in the complex boost eigenvalue plane will give rise to exponentials
modified by powers of the rapidity.  It is also directly connected to the large boost behavior, in a physically
transparent manner.

The purpose of the rest of this paper is to try to elucidate this behavior from a field theoretic point of view.
We will, for the present, restrict attention to conformal field theory. In our first section we will rewrite our
definition of the Regge limit for the four point function of general primary fields. We will see that it
corresponds to taking one cross ratio to zero with the other held fixed. We contrast this with the definition in
\cite{port}.  We explain why {\it any} conformal partial wave expansion of the four point function, corresponding
to the classification of states in terms of local operators acting on the vacuum, fails to capture the Regge
limit.  This mirrors the traditional failure of ordinary partial wave expansions for scattering amplitudes.

In Section 3, our off-shell definition of the Regge limit allows us to study first the behavior of composite
operator correlation functions in free field theory. We will find that Regge limits of four point functions have a
universal character, and that the behavior of the Regge limit for general states $| S_{e,o} \rangle $ is
determined by the matrix element of the large boost between states created by nearly light-like bilocals in the
free fields\footnote{In interacting non-abelian gauge theory, the definition of these bilocal operators includes a
light-like Wilson line.}. These indeed have the expected pure Regge behavior.  The Regge limit of higher point
functions is more complicated and appears to depend on the choice of operator at the point we decide to define as
the origin of the boost.

In section 4 we study lines of fixed points that pass through a free field point and show that close to the free
field point one can understand Regge behavior in terms of BFKL-like kernels, which depend on four space-time
points. The kernels are defined as matrix elements of the boost operator between a pair of bi-local states.
Finally, section 5 is devoted to a description of the calculation of the Regge limit of four point functions in
strongly coupled ${\cal N} = 4$ gauge theory, using the AdS/CFT correspondence.  This is a review of the work of
\cite{bpst}.  Comparing their work with a recent paper by Hofman and Maldacena\cite{mh}, we argue that dominance
of the Regge limit by the exchange of a bi-local operator in the t channel is valid at strong coupling.

\section{Conformal 4 point functions}

We will follow the conventions of Dolan and Osborn\cite{do}, and consider the four point function of scalar
operators $\phi_i $ in a four dimensional CFT.  Define $\Delta^{\pm}_{ij} \equiv \Delta_i \pm \Delta_j $, where
$\Delta_i$ is the conformal dimension of $\phi_i $.  The conformal group is isomorphic to $SO(2,4)$ and one can
realize Minkowski space as the projective light cone $$\eta^2 \equiv g_{AB} \eta^A \eta^B = 0 ; \ \ \ \eta^A \sim
\lambda \eta^A ,$$ in a six dimensional space with metric $g_{AB} = {\rm diag}\ (-1,1,1,1,1,-1) .$  The conformal
properties of Green's functions are most easily summarized by viewing the fields as functions of $\eta$, with
their dimensions $\Delta_i$ specifying how they transform under the projective equivalence: $\phi (\lambda\eta ) =
\lambda^{- \Delta} \phi( \eta )$.

It is important to understand that the Green's functions can be viewed as expectation values in two quite
different Hilbert spaces, which are related to different parameterizations of the projective light cone. These two
formulations of the theory are {\it not} unitarily equivalent to each other.  On the one hand, if we write $$\eta
= (\cos\tau , \hat{N}, \sin\tau),$$ with $\hat{N}^2 = 1$, we get a formulation of the theory on the
manifold\footnote{$\tau$ is not a periodic variable if the dimensions in the CFT are not quantized, as is the
generic case.} $R \times S^3$. On the other hand, we can write, using light front coordinates in the $45$ plane,
$$\eta = (x^{\mu} , \frac{1}{2} x^2, 1), $$ where we have used the projective invariance to set the second light
front coordinate to $1$. This gives a formulation of the Green's functions in Minkowski space.

The Hilbert space corresponding to the $R \times S^3$ parametrization carries a unitary representation of the full
$SO(2,4)$ conformal group.  In that representation, a Minkowski boost generator $J_{0i}$ in an $SO(1,3)$ subgroup
is conjugate to the dilatation generator $D = J_{45}.$  On the other hand, in the Minkowski formulation, only the
Poincar\'e $\times$ Dilatation subgroup, whose generators are $J_{0i}, J_{ij}, J_{\mu , 4 + 5}, J_{45}$, is
unitarily implemented. The operators of finite special conformal transformations do not act unitarily on the
Hilbert space, despite the fact that their generators $K_{\mu} = J_{\mu , 4 - 5}$, do act as formally Hermitian
operators. In this formulation, $D$ is not conjugate to a Minkowski boost. $SO(2,4)$ acts as an automorphism group
of the operator algebra of the theory, only part of which is implemented unitarily.

The two Hilbert spaces are related by a non-isometric map which implements the singular conformal transformation
analogous to the stereographic projection of a two-sphere on a plane. In relating Green's functions in the two
formulations, one must also take into account the conformal anomaly. Both formulations allow Euclidean
continuation, but the Euclidean theories are also inequivalent. The first describes thermal expectation values for
the operator $K_0 + P_0$, while the second constructs the statistical mechanics of the operator $P_0$, acting on a
different Hilbert space.  The Regge limit we will define, makes sense only in the ${\cal P}\times D$ Hilbert
space, in which boosts and dilatation operators have different spectra.

Dolan and Osborn describe the four point function in language appropriate to a partial wave expansion in a
particular channel. In our case, the appropriate channel is the t-channel, even though we will see that the
partial wave expansion is not relevant to the Regge limit.  Thus we write 
\begin{equation}\label{fourpnt} \langle \phi_1 (\eta_1 ) \phi_2
(\eta_2 )\phi_3 (\eta_3 )\phi_4 (\eta_4 ) \rangle = \eta_{13}^{ - \ha \Delta_{13}^+} \eta_{24}^{ - \ha
\Delta_{24}^+} \left(\frac{\eta_{34}}{\eta_{14}}\right)^{\ha \Delta_{13}^-}
\left(\frac{\eta_{14}}{\eta_{12}}\right)^{\ha \Delta_{24}^-} F(u,v),
\end{equation}
 where $\eta_{ij} \equiv \eta_i \cdot
\eta_j$, and $u,v$ are the conformally invariant cross ratios

$$ u = \frac{\eta_{13}\eta_{24}}{\eta_{12}\eta_{34}}, \ \ \ v = \frac{\eta_{14}\eta_{23}}{\eta_{12}\eta_{34}} .$$

\begin{figure}[h] \begin{center} \includegraphics[width=6cm]{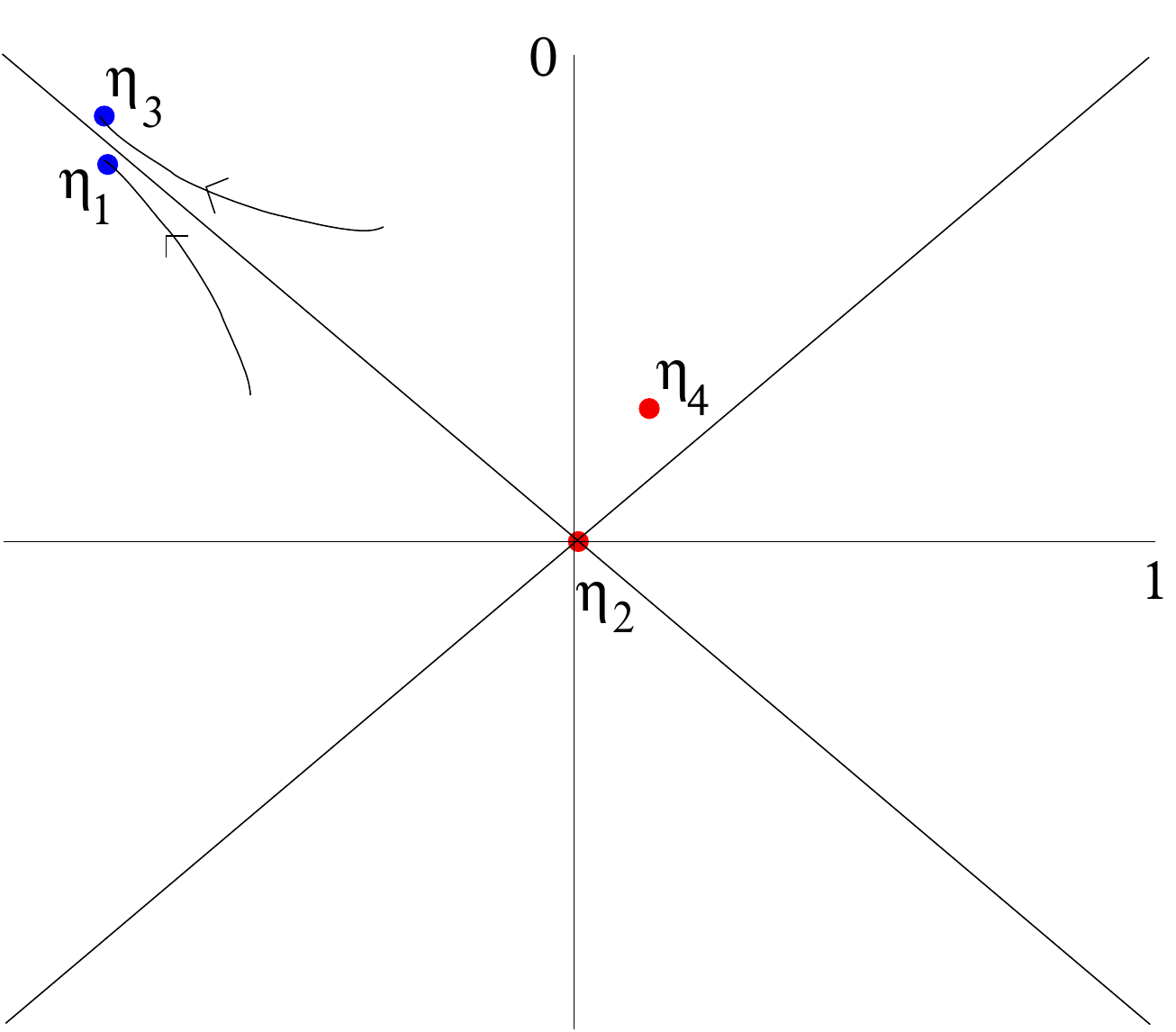} \end{center} \caption{Effect of the
boost in the $01$ plane} \label{01plane} \end{figure}

Our definition of the Regge limit is a boost of $\eta_1$ and $\eta_3$ with rapidity $\omega \rightarrow\infty$ in
(say) the $1$ direction. The boost leaves $\eta_2$ invariant. The space time causal structure in the $01$ plane is
displayed in figure \ref{01plane}. In this limit the scalar products $\eta_{14}$ and $\eta_{34}$ scale like
$e^{\omega}$, while all others remain constant. The cross ratio $u$ scales like $e^{-\omega}$, while $v$ is fixed.
The prefactor in the four point function scales like $e^{\ha\omega\Delta_{24}^{-}}$, which is a constant when
vacuum quantum numbers are exchanged in the $t$ channel.

For the four point function, our definition of the Regge limit of the four point function, is thus to take one
cross ratio to zero, with the other fixed.  This is analogous to the standard prescription for scattering
amplitudes, where $s$ is taken to $\infty$ at fixed $t$, but is expressed in terms of conformally invariant
quantities.  Note that this is different from the prescription of \cite{port} where both cross ratios are taken to
extreme values, with the variable $\frac{z}{\bar{z}}$ fixed. We believe that this difference stems from the intuition the authors of \cite{port}
gained from studying the eikonal approximation to scattering in bulk AdS space. This regime does not exist for a general CFT, and requires the large gap in anomalous dimensions for most operators in the theory, which allows one to have a quasi-local bulk description.  In particular, the eikonal approximation requires the bulk impact parameter to be of order the AdS radius or larger, which is dual to taking some sort of UV limit in the CFT. Thus, we believe that the Regge regime defined in \cite{port} is in some ways analogous to the BFKL hard pomeron regime in QCD (and indeed, in weakly coupled ${\cal N} = 4$ SYM the two approximations agree).  We have shown above that this limit corresponds to taking a relative boost which leaves none of points in the correlator fixed.
It gives different scaling exponents than the corresponding momentum space Regge limit, and in free field theory, these differences depend on the underlying field content, the Feynman diagram, and the choice of composite operators. For simple free field theories the four point function~\ref{fourpnt} is a sum of terms of the form ${u^n v^m}$ with $n$ positive\cite{do} therefore the limit considered in \cite{port} can be obtained from the one we propose by taking $v\rightarrow 1$. Arguably in the interacting case the situation could be more complex and the two limits $u\rightarrow 0$ and $v\rightarrow 1$ not commute. 

In Euclidean CFT the limit of one vanishing cross ratio is the limit controlled by the OPE or conformal partial
wave expansion. The authors of \cite{do} have determined the form of the conformal partial waves exactly.  For
spin $L$ and dimension $D$ exchange, they are given by

$$\frac{1}{x - z} [x^{\lambda_1 + 1} z^{\lambda_2} F(\lambda_1 + a, \lambda_1 + b , 2\lambda_1 ; x)F(\lambda_2 + a
-1, \lambda_2 + b - 1 , 2\lambda_2  - 2; z) - (x \rightarrow z)].$$ Here the $F$'s are the usual regular
hypergeometric series around the origin and $$a =-\ha \Delta_{13}^-,\;  b = \ha \Delta_{24}^-,\;\lambda_1={1\over
2}(D+L),\;\lambda_2={1\over 2}(D-L),\; u = xz,\;v = (1 - x)(1 - z)$$.

However, as shown very carefully in \cite{port}, for the causal relations implied by the Regge limit, one must
analytically continue these functions, picking up terms proportional to the solutions of the hypergeometric
equations that are singular at the origin.  When that is done properly one sees that higher terms in the partial
wave expansion become more and more singular in the Regge limit. This is exactly analogous to what happens in
scattering theory, where higher spin exchange in the $t$ channel is more and more singular in the Regge limit.

The authors of \cite{port} propose to deal with this by methods analogous to those of traditional Regge theory.
They rewrite the partial wave expansion as an integral, using the Sommerfeld-Watson transformation, and then {\it
hypothesize} that the leading singularity in complex angular momentum is a pole.  We prefer to use the fact that
we have exactly soluble CFTs, to guess what the general behavior is. Indeed, as we shall see, free field theory
already has interesting Regge behavior, and this will enable us to formulate a general conjecture.

\section{Regge behavior in free CFT}

Free CFTs have no scattering states, and even free massive theories have no scattering amplitudes. Nonetheless,
our definition of the Regge limit for correlation functions of composite operators is applicable to free CFTs, and
gives interesting results.  The computation of any such composite operator Green functions, in any free CFT,
reduces to a complicated Feynman diagram.  As usual, we restrict attention to connected Green functions. We will
find the analysis simplest in terms of the Lorentz frame in which all the odd points are boosted to the right,
while all the even points are un-boosted. We choose the conjugacy class of the boost in the Poincar\'e group, by
insisting that it leaves the point $x_2$ invariant. We will concentrate on the case of four point functions of
composite scalar operators. Results are similar for operators with Lorentz indices, but the actual Regge behavior
depends on how the indices are chosen on each side of the t-channel. For higher point functions, the results
depend on which operator is chosen to sit at the invariant point $x_2$ and a more refined analysis is needed.

We will be discussing theories in four dimensions, containing free fields of spin $s = 0, \frac{1}{2}, 1$.  For
spin $1$ we refer always to the gauge invariant field strength. Under boosts, these fields break up into $2s + 1$
representations, transforming as $e^{m \omega}$, with $m = -s, -s + 1, \ldots ,s$.

The propagators in the diagram for any such Green function can be divided into three classes:

\begin{itemize}

\item I. Propagators connecting two even points

\item II. Propagators connecting two odd points, or connecting an odd point to $x_2$.

\item III. Propagators from an odd point, to $x_4$.

\end{itemize}

Propagators of class I are independent of $\omega$ for all values of $(s,m)$.  Propagators of Class II behave
like

$$e^{(m_1 + m_2)\omega },$$ while propagators of class III behave like $$e^{((m_1 + m_2) - 2s -1)\omega}.$$

To proceed, we must recall that Regge behavior depends on the quantum numbers exchanged in the t-channel. Free
field theory has an infinite number of conservation laws.  We define vacuum exchange by insisting that the two
pairs of operators separated by the boost, each have a vacuum expectation value.  This means that the number (and
kind) of free fields in each of the composite operators in a pair is the same (recall that no self contractions
are allowed by the definition of normal ordered operators). Thus for example the four point function $$\langle T\
\phi^2 (x_1 ) \phi^4 (x_3) \phi^2 (x_2) \phi^4 (x_4) \rangle ,$$ does {\it not} correspond to a vacuum exchange,
because the two point function of $\phi^2$ and $\phi^4$ vanishes. It then follows that the number $b$ of
propagators connecting $x_1$ to $x_4$ and $x_2$ to $x_3$ must be the same. This is true as well for the number $c$
of propagators
 connecting $x_1$ to $x_2$ and $x_3$ to $x_4$.

Due to the scalar nature of the composite operators the dominant contribution of each graph in the large $\omega$
limit is at most $e^{-\omega(b+c)}$. It then follows that the Regge limit is dominated by those graphs with
$b+c=1$ that is to say where just two lines are exchanged across the $t$ channel.

It is also clear that the Regge behavior is somewhat universal. All that really matters is what is going on at the
vertices of the diagram from which the relevant cross-channel propagators emanate. This can be rephrased in a way
which lends itself to far reaching generalization. In the introduction, we expressed the Regge limit as the limit
of the matrix element of a boost operator between two states, $| S_{e,o} \rangle $. Our free field analysis shows
us that the limit is dominated by the overlap of those states with certain {\it light-like multi-local operators}.
We use this phrase to refer to a product of local operators whose positions in the boost plane are confined to one
of the two light rays defined by the boost.  In fact, we believe that all that is relevant are bi-local
operators.

In the Regge limit of four point functions, all that appears in the t-channel analysis is states created by
light-like bi-locals  The set of all states create by acting on the vacuum with light-like bi-local operators,
whose component local operators are arbitrary primary fields, is over-complete. The analysis above shows that only
the component of a state which has an overlap with the light-like bi-local whose constituents are elementary
fields (for $s = 1$ this means the gauge invariant field strength), is relevant in the Regge regime.  For models
with various values of $s$ it is the free fields with highest $s$, which dominate in the Regge limit.

For non-abelian gauge groups, the field strength bi-local is not gauge invariant when the coupling is different
from zero.  In this case it will be replaced at finite coupling by $$ {\rm Tr}\ [U^{\dagger} (x,y) F (x) U(x,y)
F(y)],$$ where $U(x,y)$ is the Wilson line in the fundamental representation. When the gauge potentials are
normalized so that the free field Lagrangian is $g$ independent, this formula reduces to the sum of abelian field
strength two point functions when $g \rightarrow 0$.  We anticipate that these gauge invariant operators will be
relevant both in asymptotically free theories and in theories with a line of fixed points passing through the
origin.

\section{Bi-local operators and the Low-Nussinov-BFKL ansatz}

For vacuum quantum number (pomeron) exchange, in free QCD, the leading Regge behavior is dominated by the states

$$ {\rm Tr}\ F_{+ i} (x^+ , {\bf x}) U(x,y) F_{- j} (y^{+ j}, {\bf y})U^{\dagger} (x,y) | 0 \rangle ,$$ where
${\bf x}$ is a transverse vector. In the interacting theory, the field strength should be the full gauge covariant
non-abelian field strength and $U$ is the null Wilson line\footnote{We note that Balitsky\cite{balitsky} has proposed that the Regge limit in QCD can be
determined by the behavior of null Wilson lines.}.  When the coupling is zero, $U = 1$, and $F$ is just
the Maxwell field strength.   The dominant diagrams involve two gluons exchanged between bi-local operators on
oppositely directed null rays.   This is the Low-Nussinov model for the pomeron in QCD\cite{ln}.

If, as in the case of ${\cal N} = $ SYM theory, we have a line of fixed points passing through the origin, we
should expect that at small coupling the same bi-local operator dominates the Regge limit, but the two gluon
exchange graph is decorated with interaction corrections.  Thus, one must extract from the four gluon amplitude,
the dominant set of graphs in the Regge limit.   We can write a Bethe-Salpeter equation

$$A(x_1, x_2, x_3, x_4) = K (x_1, x_2, x_3, x_4) + \int\ dy dz K(x_1, y, x_3 , z) A(y, x_2, z, x_4 ), $$ where the
kernel $K$ is two particle irreducible in the $t$ channel.
\begin{figure}[h] \begin{center} \includegraphics[width=10cm]{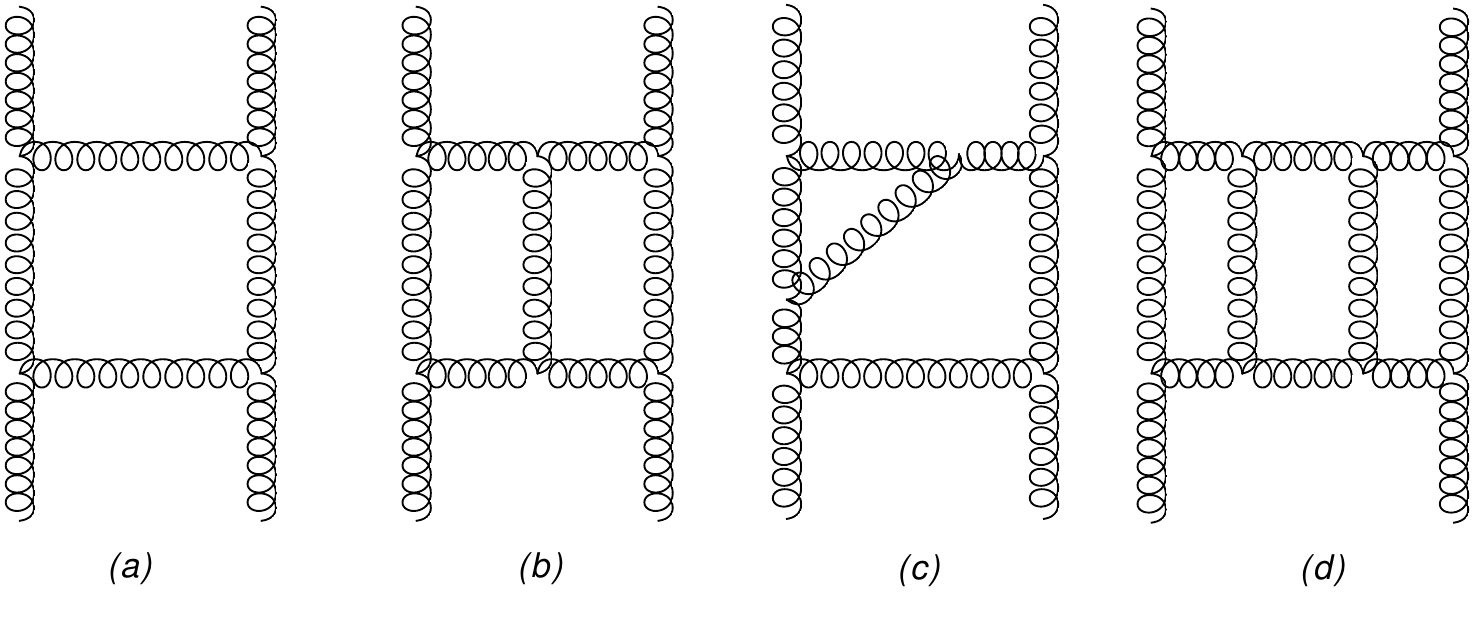} \end{center} \caption{(a) is the dominant graph for the kernel $K(x_1,x_2,x_3,x_4)$, (b) (c) and (d) are subdominant ones.} \label{gluonkernel}
\end{figure}

  If we look at a general graph contributing to the
amplitude, and choose the Lorentz frame where the odd points are boosted and the even points left at rest, it is
easy to see that all $\omega$ dependence comes from the {\it last rung of the ladder} where the rest of the graph
connects to the points $x_1$ and $x_3$.  The Regge limit is dominated by graphs with a minimal number of lines in
this last rung.   Figure \ref{gluonkernel} shows the dominant graph and some sub-dominant ones.  It is clear that the dominant
Regge behavior is given by the solution of the Bethe-Salpeter equation with an approximate kernel.
\cite{symbfkl} have done a careful analysis of the dominant contributions to the kernel, and established the
leading order Regge behavior in the weakly coupled SYM theory. They find a $t$ dependent Regge exponent, as
expected.  We do not have anything to add to the quantitative part of their analysis, but we hope that our
considerations have made it clear why the BFKL ansatz determines the leading Regge behavior at weak coupling.

Note that at present we have nothing to say about the subtleties of the BFKL analysis in asymptotically free
theories. One would hope to return to this point once a more complete understanding of conformal field theories is
achieved.

\section{The transverse conformal group}

The little group of a light-like line is the product of a longitudinal conformal group $SL(2,R)$ and a transverse
conformal group $SL(2,C)$ (in four dimensions).  Scattering in the Regge limit involves two oppositely directed
light-like lines, and is thus invariant only under the transverse conformal group.  That is, the Bethe-Salpeter
kernel of the previous section should be covariant under the transverse conformal group and we can use group
theory to reduce the equation. This has already been exploited in the literature on the BFKL limit. In this
section, we give a brief discussion of the transformation properties of states created by null bi-locals under the
transverse conformal group.

Consider bilocal operators built from quasi-primary scalar operators of conformal dimensions $\Delta$ \footnote{we only consider two operators of the same dimension because we are eventually interested in the vacuum exchange channel for the four point function.}.
\begin{equation}
O^2_{1,2}(f)=\int d^4 x d^4 y f(x,x')O_1(x)O_2(x')
\end{equation}
Under a conformal transformation $x\rightarrow y$ $O(x)\rightarrow O'(y)=|{\partial y\over \partial x}|^{-{\Delta \over 4}}O(x) $ these transform to:
\begin{equation}
\int d^4 x d^4 x' \left|{\partial y'\over \partial x'}\right|^{1-{\Delta\over 4}}\left|{\partial y \over \partial x}\right|^{1-{\Delta\over 4}}f( y(x),y'(x'))O_1(x)O_2(x')
\end{equation}
For $f(x,y)$ proportional to $\delta(x^-)\delta(x'^-)$ and introducing complex coordinates $(z,z')$ in the transverse plane we consider the operators:
\begin{eqnarray}
O^2_{1,2}(f)=\int_{0}^{\infty} d x^+ d x'^+\int d^2z d^2 z' f(x^+,x'^+,z,z')O_1(x^+,0,z)O_2(x'^+,0,z)
\end{eqnarray}
The two points we integrate over are spacelike separated, unlike the causal relation considered in section 2 and illustrated in figure \ref{01plane}. However we are interested in the behavior of the bilocals under a large boost in the $(+,-)$ plane under which the $x^+$ separation of the two points increases while the $x^-$ separation shrinks. In this limit a very small shift in the $x^-$ coordinates of the two points defining the bilocal restores the previously considered causal relation.

The action of the transverse conformal group and boost in the $(+,-)$ plane is:
\begin{itemize}
\item{Under the boost $f(x^+,x'^+,z,z')\rightarrow \lambda^{2}f(\lambda x^+,\lambda x'^+,z,z')$}
\item{Under translations $z\rightarrow z+a$ in the transverse plane $f(x^+,x'^+,z,z')\rightarrow f(x^+,x'^+,z+a,z'+a)$}
\item{Under $z\rightarrow \lambda z,\;\;z\in C$ corresponding to dilatations and rotations in the transverse plane $f(x^+,x'^+,z,z')\rightarrow|\lambda|^{6-2\Delta}f(|\lambda|x^+,|\lambda|x'^+,\lambda z,\lambda z')  $}
\item{Under $z\rightarrow z(1-a z)^{-1}, x^+\rightarrow x^+ |1-az|^{-2}$ which are the special conformal transformations in the transverse $SL(2,C)$:
    $$f(x^+,x'^+,z,z')\rightarrow |1-a z|^{2\Delta-6}|1-az'|^{2\Delta-6} f(x^+ |1-az|^{-2},x'^+ |1-az'|^{-2},z(1-a z)^{-1},z'(1-a z')^{-1})$$}
\end{itemize}
The infinitesimal form of these transformations are starting with the boost
\begin{equation}
x^{+}\partial_{x^+}+y^{+}\partial_{y^+}+2
\end{equation}
while the transverse $SL(2,C)$ act as
\begin{eqnarray}
&&\partial_z+\partial_z'\\
&&z\partial_z+z'\partial_{z'}+(1/2)(x^+\partial_{x^+}+y^+\partial_{y^+})+(1/2)(6-2\Delta) \\
&&z^2 \partial_z+z'^2 \partial_{z'}+z x^+ \partial_{x^+}+z' y^+\partial_{y^+}+ (z+z')(3-\Delta)
\end{eqnarray}
and the respective anti-holomorphic ones.
From this we obtain the expression for $L^2$
\begin{eqnarray}
&&-(z-z')^2\partial_z \partial_{z'}+\eta'(z-z')\partial_z+\eta(z'-z)\partial_z'+(1/4)(\eta+\eta')(\eta+\eta'-2)\\
&&\eta'=(3-\Delta+y^+\partial_{y^+}),\;\;\;\eta=(3-\Delta+x^+\partial_{x^+})
\end{eqnarray}
We can now restrict the form of the function $f(x^+,y^+,z,z',\bar{z},\bar{z}')$ to be:
\begin{equation}
(x^+)^{i \gamma_1-d_1}(y^+)^{i \gamma_2-d_{2}}f(z,z')g(\bar{z},\bar{z}')
\end{equation}
In the following we set $d_1=d_2=1$ that is $\eta=2-\Delta+i\gamma_1,\;\;\eta'=2-\Delta+i\gamma_2$.

Over these functions we can diagonalize $L^2$ getting the following
\begin{eqnarray}
&&f(z,z')=(z-z_0)^{\alpha-\eta}(z'-z_0)^{\alpha-\eta'}(z-z')^{-\alpha}\\
&&g(\bar{z},\bar{z}')=(\bar{z}-\bar{z}_0)^{\tilde{\alpha}-\eta}(\bar{z}'-\bar{z}_0)^{\tilde{\alpha}-\eta'}(z-z')^{-\tilde{\alpha}}\\
&&(\alpha-{1\over 2}(\eta+\eta'))(\alpha+1-{1\over 2}(\eta+\eta'))=h(h-1)\\
&&(\tilde{\alpha}-{1\over 2}(\eta+\eta'))(\tilde{\alpha}+1-{1\over 2}(\eta+\eta'))=\tilde{h}(\tilde{h}-1)
\end{eqnarray}
For what concerns $SL(2,C)$ the $d_i=1$  just shift $\Delta\rightarrow \Delta+1$  so in the following we will work with the shifted value unless otherwise noted.
As we want a unitary representation of $SL(2,C)$ the parameter $h$ has to be chosen $h=((n+1)/2-i \nu)$ where $n\in Z$ and $\nu\in R$. The two solutions for $\alpha$ are then $\alpha=h-1+(1/2)(\eta+\eta')$ and $\alpha=-h+(1/2)(\eta+\eta')$. They are related by $\nu\rightarrow -\nu$ and $n\rightarrow -n$. The corresponding two 
$\tilde{\alpha}=-\bar{h}+(1/2)(\eta+\eta')$ and $\tilde{\alpha}=\bar{h}-1+(1/2)(\eta+\eta')$ respectively. It follows that $\alpha$ is related to $\tilde{\alpha}$ by $n\rightarrow -n$.
The form of the functions $f(z,z')$ is that of a three point function for an operator of weight $\eta$ at $z$ one of weight $\eta'$ at $z'$ and one of weight $1-h$ at $z_0$.

Denote $F^{n,\nu}(z-z_0,z'-z_0)=f(z-z_0,z'-z_0)g(\bar{z}-\bar{z}_0,\bar{z}'-\bar{z}_0)$. It is useful to define the following transform\cite{BFKL}:
\begin{eqnarray}
&&F_{q,\bar{q}}^{n,\nu}(\rho)={c_{n,\nu}\over 2\pi} \int d^2 z e^{{i\over 2}(\bar{q}z+q\bar{z})}F^{n,\nu}\left(z+{\rho\over 2},z-{\rho\over 2}\right)\nonumber \\=&&{c_{n,\nu}\over 2 \pi}(\rho\bar{\rho})^{{1\over 2}(\alpha+\tilde{\alpha})-\eta -\eta'+1}\left(\rho\over \bar{\rho}\right)^{{n\over 2}} \int ^2 z e^{{i\over 2}(\bar{q}\rho z+q\bar{\rho}\bar{z})}F^{n,\nu}\left(z+{1\over 2},z-{1\over 2}\right)\nonumber\\
&&F^{n,\nu}(z-z_0,z'-z_0)={1\over 2 \pi c_{n,\nu}} \int d^2 q e^{-{i\over 2}(\bar{q}(z+z'-2z_0)/2+c.c.)}F_{q,\bar{q}}^{n,\nu}(z-z')
\end{eqnarray}
where $c_{n,\nu}$ will be chosen shortly.
The behavior of $F^{n,\nu}_{q,\bar{q}}(\rho)$ for $\rho\bar{q}\rightarrow 0$ comes from two regions in the integral: either $z$ is of $O(1)$ and the exponential can be expanded or $z\sim {1\over \rho \bar{q}}$ and we can neglect the ${1\over 2}$ in $F^{n,\nu}\left(z+{1\over 2},z-{1\over 2}\right)$ therefore:
\begin{eqnarray}
&&\lim_{\bar{q}\rho\rightarrow 0}F_{q,\bar{q}}^{n,\nu}(\rho)={c_{n,\nu}\over 2 \pi}(\rho\bar{\rho})^{-i \nu-{1\over 2}(\eta +\eta')+{1\over 2}}\left(\rho\over \bar{\rho}\right)^{{n\over 2}}\cdot\nonumber \\&&\cdot\left( \int d^2 z F^{n,\nu}\left(z+{1\over 2},z-{1\over 2}\right)+2^{-4i \nu}({\bar{q}\rho q\bar{\rho}})^{2 i \nu}\left(\bar{q}\rho\over q \bar{\rho}\right)^{-{n}}\int d^2 z e^{{i\over 2}( z+\bar{z})}(z\bar z)^{\alpha+\tilde{\alpha}-\eta-\eta'}\left(z\over \bar{z}\right)^{n}\right)\nonumber
\end{eqnarray}
Now  choose the coefficient $c_{n,\nu}$ such that
\begin{eqnarray}
&&c_{n,\nu}^{-1}={1\over 2\pi} \int d^2 z F^{n,\nu}\left(z+{1\over 2},z-{1\over 2}\right)= \nonumber\\= &&{ (-1)^n\over 2( 1-2\bar{h})} {\Gamma(1-\bar{h}-{i\over 2}(\gamma_1-\gamma_2))\Gamma(1-\bar{h}+{i\over2}(\gamma_1-\gamma_2))\Gamma(1-2h)\over \Gamma(1-{h}-{i\over 2}(\gamma_1-\gamma_2))\Gamma(1-{h}+{i\over2}(\gamma_1-\gamma_2))\Gamma(1-2\bar{h})}
\end{eqnarray}
And get
\begin{eqnarray}
\lim_{\bar{q}\rho\rightarrow 0}F_{q,\bar{q}}^{n,\nu}(\rho)=(\rho\bar{\rho})^{-i \nu-{1\over 2}(\eta +\eta')+{1\over 2}}\left(\rho\over \bar{\rho}\right)^{{n\over 2}}\cdot\left( 1+\xi(n,\nu)2^{-4i \nu}({\bar{q}\rho q\bar{\rho}})^{2 i \nu}\left(\bar{q}\rho\over q \bar{\rho}\right)^{-{n}}\right)
\end{eqnarray}
Where $\xi(n,\nu)$ is equal to:
\begin{equation}
\left({n+i\nu\over n-i\nu}\right){\Gamma(1-{h}-{i\over 2}(\gamma_1-\gamma_2))\Gamma(1-{h}+{i\over2}(\gamma_1-\gamma_2))\Gamma(1-2\bar{h})^2\over \Gamma(1-\bar{h}-{i\over 2}(\gamma_1-\gamma_2))\Gamma(1-\bar{h}+{i\over2}(\gamma_1-\gamma_2))\Gamma(1-2h)^2}
\end{equation}
Note that $\xi(n,\nu)$ is of unit norm. Also the following relation holds $\xi(-n,-\nu)\xi(n,\nu)=1$

Next from conformal invariance the following relation holds:
\begin{eqnarray}
\label{inv}
F_{q,\bar{q}}^{-n,-\nu}(\rho)=K(n,\nu)\left({2\over \bar{q}}\right)^{1-2h}\left({2\over q}\right)^{2\bar{h}-1}F_{q,\bar{q}}^{n,\nu}(\rho)\nonumber
\end{eqnarray}
The coefficient $K(n,\nu)$ can be obtained by looking at the limit for $\bar{q}\rho\rightarrow 0$.
$$
K(n,\nu)=\xi^{-1}(n,\nu)=\xi(-n,-\nu)
$$
In coordinate space this relation reads:
\begin{equation}
F^{-n,-\nu}(z-z_0,z'-z_0)={c_{n,\nu}\over 2\pi} \int d^2 w (z_0-w)^{-2h} (\bar{z}_0-\bar{w})^{2\bar{h}-2}F^{n,\nu}(z-w,z'-w)
\end{equation}
The orthogonality formula for the $F^{n,\nu}_{q,\bar{q}}(\rho)$ has the following form dictated by conformal invariance:
\begin{eqnarray}
&&\int d^2 \rho (\rho\bar{\rho})^{4-2\Delta}F_{q,\bar{q}}^{n,\nu}(\rho)\bar{F}_{q,\bar{q}}^{n',\nu'}(-\rho)=\nonumber\\ && =A(n,\nu)\delta_{n,n'}\delta(\nu-\nu')+B(n,\nu)\delta_{n,-n'}\delta(\nu+\nu')\left({q\over \bar{q}}\right)^{n}(q\bar{q})^{2 i \nu}
\end{eqnarray}

The coefficients can be obtained by using the limiting form of the $F_{q,\bar{q}}^{n,\nu}(\rho)$ and are:
\begin{eqnarray}
A(n,\nu)&=&(2 \pi)^2 (-1)^n\\
B(n,\nu)&=&(2\pi)^2 \xi(n,\nu)2^{-4 i \nu} (-1)^n
\end{eqnarray}
This relation in coordinate space reads
\begin{eqnarray}
&&\int d^2 z d^2 z' |z-z'|^{8-4\Delta}F^{n,\nu}(z-z_0,z'-z_0)\bar{F}^{n',\nu'}(z'-z_0',z-z_0')=\nonumber \\&&=A'(n,\nu)\delta_{n,n'}\delta(\nu-\nu')\delta^2(z_0-z_0')+B'(n,\nu)\delta_{n,-n'}\delta(\nu+\nu')(z_0-z_0')^{2h-2}(\bar{z}_0-\bar{z}_0')^{-2\bar{h}}\nonumber
\end{eqnarray}
Where the constants $A'(n,\nu),\;B'(n,\nu)$ are given by.
\begin{eqnarray}
A'(n,\nu)&=&(2 \pi)^4 {(-1)^n\over 4 (n^2+4 \nu^2)}\\
B'(n,\nu)&=&(2\pi)^4  {(-1)^n\over 4 (n^2+4 \nu^2)}{c_{-n,-\nu}\over 2\pi}
\end{eqnarray}
Finally the completeness relation reads:
\begin{eqnarray}
&&(2\pi)^4\delta^2 (z-x)\,\delta^2(z'-x')=|z-z'|^{4-2\Delta}|x-x'|^{4-2\Delta}\cdot \nonumber \\&&\cdot\sum_{n=0}^{\infty}(-1)^n \int d\nu \int d^2 q e^{-{i\over 4}(\bar{q}(z+z'-x-x')+c.c.)} F_{q,\bar{q}}^{n,\nu}(z-z')\bar{F}_{q,\bar q}^{n,\nu}(x'-x)
\end{eqnarray}
This is also readily written in coordinate space:
\begin{eqnarray}
&&(2\pi)^4\delta^2 (z-x)\,\delta^2(z'-x')=|z-z'|^{4-2\Delta}|x-x'|^{4-2\Delta}\cdot \nonumber \\&&\cdot\sum_{n=0}^{\infty}(-1)^n \int d\nu G(n,\nu) \int d^2 w F^{n,\nu}(z-w,z'-w)\bar{F}^{n,\nu}(x'-w,x-w)
\end{eqnarray}
where $G(n,\nu)=4(n^2+4\nu^2)$

The bilocal operators $O^2_{n,\nu,\gamma_1,\gamma_2}(w)$ transform under the
transverse conformal group $SL(2,C)$ as fields of weight $(-1-2i\nu+n,-1-2i\nu -n)$ and get multiplied by $e^{i(\gamma_1+\gamma_2)\chi}$ under a boost of rapidity $\chi$ in the $(+,-)$ plane. An equivalent construction can be done for bilocal operators with $x^+=x'^+=0$. Then we can write for the four point function when the first two points are in the future of the second couple $\langle O_{\Delta}(x^+,x'^+,z,z')O_{\Delta'}(x^-,x'^-,x,x')\rangle$ (and expliciting the shift in $\Delta$ due to $d=1$):
\begin{eqnarray}
&&(2\pi)^{-12} \sum_{n=0}^{\infty} \int d\nu G^2(n,\nu)\int d^2 w d^2 w' d\gamma_1 d\gamma_2 d\gamma_1' d\gamma_2' (x^+)^{-i\gamma_1}(x'^+)^{-i\gamma_2}(x^-)^{-i\gamma_1'}(x'^-)^{-i\gamma_2'} \cdot\nonumber\\&& |x-x'|^{4-4\Delta}|z-z'|^{4-4\Delta'}\bar{F}^{n,\nu}(z'-w,z-w)\bar{F}'^{n,\nu}(x'-w',x-w')\cdot \nonumber \\&&\delta(\gamma_1+\gamma_2-\gamma_1'-\gamma_2')K(n,\nu,\gamma_1,\gamma_2,\gamma_1',\gamma_2')(w-w')^{n-1-2i\nu}(\bar{w}-\bar{w}')^{-1-n-2i\nu}
\end{eqnarray}
The integrals in $w$ and $w'$ can be done using \ref{inv} and the formula in appendix A obtaining:
\begin{eqnarray}
&&-(2\pi)^{-10} \sum_{n=0}^{\infty}\int d\nu d\gamma_1 d\gamma_2 d\gamma_1' d\gamma_2' (x^+)^{-i\gamma_1}(x'^+)^{-i\gamma_2}(x^-)^{-i\gamma_1'}(x'^-)^{-i\gamma_2'} \cdot\nonumber\\&& | z'- z|^{-2\Delta-{i}(\gamma_1+\gamma_2)}|x'-x|^{- 2\Delta'-{i}(\gamma_1'+\gamma_2')} |z'-x|^{{i}(\gamma_1-\gamma_1'-\gamma_2+\gamma_2')} \cdot \nonumber \\&&\delta(\gamma_1+\gamma_2-\gamma_1'-\gamma_2')K(n,\nu,\gamma_1,\gamma_2,\gamma_1',\gamma_2')\nonumber\\
&&\Bigl(G(n,\nu) H(n,\nu,\gamma_1,\gamma_2,\gamma_1',\gamma_2',y,\bar y)+{c_{-n,-\nu}^2\over G(n,\nu)}H(-n,-\nu,\gamma_1,\gamma_2,\gamma_1',\gamma_2',y,\bar y)\Bigr)
\end{eqnarray}
where $y={(z'-z)(x'-x)\over (z'-x')(z-x)}$ and
\begin{eqnarray}
&&H(n,\nu,\gamma_1,\gamma_2,\gamma_1',\gamma_2',y,\bar y)=\cr &&=y^{{1\over 2}(1+n)-i\nu}\bar{y}^{{1\over 2}(1-n)-i\nu}{_2}F_1(a,b;c;y){_2}F_1(\tilde{a},\tilde{b};\tilde{c};\bar{y})
\end{eqnarray}
with:
\begin{eqnarray}
a&=&h +{i\over 2}(\gamma_1-\gamma_2),\;\;\tilde{a}=1-\bar{h} +{i\over 2}(\gamma_1-\gamma_2)\cr
b&=&h+{i\over 2}(\gamma_2'-\gamma_1'),\;\;\tilde{b}=1-\bar{h}+{i\over 2}(\gamma_2'-\gamma_1')\cr
c&=&2h,\;\;\;\;\;\;\;\;\;\;\;\;\;\;\;\;\;\;\;\;\;\tilde{c}=2-2\bar{h}
\end{eqnarray}

This consists in writing the four point function as an integral over partial waves in the transverse plane with weights $(-1-2i\nu+n,-1-2i\nu -n)$. The behavior of the four point function under the Regge limit is governed by the closest singularity of $K(n,\nu,\gamma_1,\gamma_2,\gamma_1',\gamma_2')$ to the $\gamma$'s real axis.

For free fields the four point function only depends on integer powers of $x^+,x^-,x'^+,x'^-$ therefore the function $K(n,\nu,\gamma_1,\gamma_2,\gamma_1',\gamma_2')$ must have poles for integer imaginary values of the $\gamma$'s. Also in this case the only partial waves that contribute to the sum have integer spin and dimension\cite{do} therefore there will be poles for imaginary values of $\nu$.

\section{The strong coupling limit}

Our interest in the Regge limit of conformal field theory, stems from the seminal paper of BPST\cite{bpst}, who
used AdS/CFT to calculate the Regge limit of ``scattering amplitudes".  Their method was to expand around the flat
space limit of tree level string scattering amplitudes, where Regge behavior has been understood for decades.
Unfortunately this method obscures the actual conformal field theory behavior. ``Scattering amplitudes" in AdS
space are really boundary CFT correlators in Poincar\'e coordinates. They depend on cross ratios of boundary points,
rather than flat momentum space invariants.  There is a translation between the two, but the translation exercise
was not carried out in \cite{bpst}.

We will therefore proceed by indirection.  We briefly recapitulate the beautiful analysis of Regge behavior by
BPST, which uses a {\it Pomeron vertex operator}.   We then recall a paper of Maldacena and Hoffman\cite{mh},
which shows that such non-local vertex operators in light cone gauge string theory, correspond to the kind of null
bi-local boundary operators that we discussed above in the weak coupling limit.

BPST begin by analyzing the Regge limit of tachyon scattering in closed bosonic string theory in flat space.  They
argue that this limit is dominated by the world sheet operator product expansion in the $t$ channel, but that one
must keep a selected set of higher order terms in the OPE, corresponding to all of the states on the leading Regge
trajectory, rather than just the lowest lying one. The requisite formula is

$$ e^{ i p_1 X (w)} e^{- i p_3 X(0)} \sim |w|^{- 4 - \alpha^{\prime} t} e^{ i (p_1 - p_3) X(0) + i(w\partial +
\bar{w}\bar{\partial} ) p_1 X(0) } .$$ The point is that $|w| \sim s^{-1}$ dominates the integral, and the factors
of $p_1$ in the second term compensate for the extra powers of $w$.  One can formally do the $d^2 w$ integral, and
write the integral over the world sheet of the RHS as

$$ \Pi \left(1 + \alpha^{\prime} \frac{t}{4}\right) e^{ i (p_1 - p_3) X(0)} [p_1 \partial X p_1 \bar{\partial} X
]^{1 + \alpha^{\prime} \frac{t}{4}} ,$$ where $$\Pi (v) =  2\pi \frac{\Gamma ( - v)}{\Gamma (1 + v)} e^{ - i\pi
v}.$$

In actual fact, up to a constant, this derivation is the same for any pair of physical vertex operators.  The
point is that all physical vertex operators have the same dimension and the factor $e^{ i (p_1 - p_3) X(0)}$ is
universal. This determines the power of $|w|$ in the OPE.  The constant, which does depend on which vertex
operators are involved, represents the coupling of the boosted string states to the pomeron Regge trajectory.
Indeed, BPST prove a general formula for scattering of any string states in the Regge regime where two sets of
vertex operators ${\cal W}_{L,R}$ are separated by a large relative boost with rapidity $\omega$.  The formula is

$$ A_{{\cal W}_L {\cal W}_R} \sim \Pi (v) e^{2yv} \langle {\cal W}_{R0} {\cal V}_P^- \rangle \langle {\cal V}_P^+
{\cal W}_{L0} \rangle ,$$ where $v = 1 + \frac{\alpha^{\prime} t}{4} $, $\Pi$ is the function defined above,
${\cal W}_{L,0}, {\cal W}_{R,0} $ are the vertex operators describing the systems boosted to the left or right
respectively, {\it but in their rest frames}, and

$${\cal V}_P^{\pm} = \left(2 \frac{\partial X^{\pm} \bar{\partial} X^{\pm}}{\alpha^{\prime}} \right)^v e^{\mp i
(p_1 - p_3) X} . $$ Thus, scattering in the Regge limit in tree level bosonic string theory is described by the
coupling of the scattering systems to the {\it Pomeron vertex operators}, ${\cal V}_P^{\pm} $. Similar formulae
hold for the superstring, in the so called 0-picture, where everything is world sheet supersymmetrized in the
obvious way, and the definition of $v$ is shifted to eliminate the tachyon poles in the $\Gamma $ functions.
Details can be found in \cite{bpst}.

To apply these results to the Regge behavior of strongly coupled Super Yang Mills theory, one must generalize them
to weakly curved space.  The key point is that $t$ can be thought of in coordinate space as the transverse
Laplacian, and we can simply make this replacement in all the formulae.  To evaluate the scattering amplitudes we
must understand the eigenfunctions and eigenvalues of this operator. Flat space vertex operators are modified by
functions of the AdS radial coordinate, which are easily computable to leading order in the world sheet loop
expansion.  Again, we refer the reader to \cite{bpst} for the details.

The key feature which is {\it not} explained in \cite{bpst} is the meaning of the Pomeron vertex operator in terms
of CFT objects. Ordinary local vertex operators correspond to the stringy generalization of non-normalizable
solutions of the linearized wave equation on $AdS$ space.  The clue to this lies in the work of Maldacena and
Hofman\cite{mh} on Conformal Collider Physics.  The basic idea in \cite{mh} is to study the correlation functions
of the stress tensor and global symmetry currents in a conformal field theory, which are produced by the injection
of a large amount of energy in the remote past.   The energy of course, ends up running out to null infinity, and
so one is interested in correlators of operators of the form

$$ {\cal E} ({\bf y}) = \int dy^-\ T_{--} (y^- , y^+ = 0, {\bf y}) ,$$ where ${\bf y}$ is a transverse two vector.
When the transverse positions of two of these operators are close together, one can do an operator product
expansion, but the right hand side contains operators which are non-local in the $y^-$ direction. Maldacena and
Hofman show that the OPE in ${\cal N} = 4$ SYM theory is dominated by a small number of bi-local operators of the
type we have studied above.

In the planar approximation, the AdS/CFT correspondence tells us that any correlation function in the SYM theory
can be calculated as a world sheet expectation value in Type IIB string theory on $AdS_5 \times S^5$. For strong
't Hooft coupling, the world sheet field theory is in its semi-classical limit. In computing the small transverse
separation limit from the world sheet theory, MH encounter a world sheet OPE, where the dominant contribution
comes from a vertex operator of the form

$$ (\partial X^+ \bar{\partial} X^+ )^{3/2} \delta (X^+ ) e^{i k X} ,$$ where $k$ is a transverse vector.  The
fractional power of a light cone coordinate again indicates non-locality along a null ray. Comparison with their
general discussion indicates that one should associate such world sheet vertex operators with null bi-local
operators in the gauge theory. It should be noted that the $3$ in the exponent is related to the Lorentz spin of
the dominant operator.  This is precisely analogous to the appearance of the ``spin of the pomeron Regge pole" in
the Pomeron vertex operators of BPST.

These parallels, combined with our weak coupling results, lead us to the following general conjecture:

\begin{itemize}

\item {\it The Regge limit in a general CFT is dominated by the exchange of a small number of null-bi-local
    operators in the channel separating the left moving from the right moving operators. In gauge theories,
    the individual local operators from which the bi-local is constructed, are not gauge invariant, and are
    accompanied by the appropriate null Wilson line.}

\end{itemize}

This conjecture is the central point of our paper.

\section{Conclusions}

We have defined a Regge limit for general correlation functions in quantum field theory, and shown how to evaluate
it for correlation functions of gauge invariant composite operators in free conformal field theory. In free field
theories of spins $\leq 1$, the Regge limit is determined in terms of the matrix elements

$$ \langle 0| \eta (x_1^+ , x_1^- = 0 , {\bf x_1} )\eta (x_3^+ , x_3^- = 0 , {\bf x_3} ) U(\omega ) \eta (0,0,{\bf
x_2} )\eta (x_4^- , x_4^+ = 0 , {\bf x_4}) | 0 \rangle .$$ $\eta$ is a gauge covariant free field.  $U(\omega )$
is a boost with rapidity $\omega$ and the light front coordinates are in the plane of that boost. In particular,
for spin $1$ $\eta$ is the Maxwell field strength. In an interacting theory which is a small perturbation of the
free theory, like ${\cal N} = 4$ SYM with weak coupling, the Maxwell tensor is replaced by the non-abelian field
strength tensor, and null Wilson lines are appended to make the bi-local operator gauge invariant.

We argued that this provided a derivation of the Low-Nussinov-BFKL prescription for the pomeron, at least for
lines of conformal SYM theories passing through the origin.  The BFKL kernel is just the leading Regge limit part
of the bi-local two point function that is two particle irreducible in the $t$ channel.  The Bethe-Salpeter
equation for the full two point function of bi-locals is covariant under the transverse conformal group $SL(2,C)$
and we provided a derivation of the $SL(2,C)$ transformation properties of null bi-locals constructed from
conformal primaries.

We also reviewed the work of BPST on Regge behavior in the strong coupling limit. Comparing it to the work of
Maldacena and Hofman, we were led to conjecture that the Regge limit is dominated by the exchange of null
bi-locals in the $t$ channel in a general CFT. One of the first things that remains to be done is to sharpen this
result in the strong coupling regime.  To do this, we have to transform the BPST result into a function of the
cross ratios, verify that their Regge limit coincides with the one defined in this paper (one cross ratio to zero
with the other fixed), and try to identify the bi-local operator that dominates in the strong coupling limit.

Other important avenues for future work include the development of a deeper understanding of states created from
the vacuum by null bi-locals. If we let the local constituents of the bi-local be arbitrary primary fields, this
is clearly an over-complete set. How can we characterize the dominant term in the Regge limit in a more general
way? Some insight may be gained by studying exactly soluble non-Gaussian CFTs in $1 + 1$ dimensions, though one
may worry that some of the results will be somewhat special.

The most important task to be accomplished is the study of how all of this generalizes to non-conformal theories
and in particular to QCD.  There is a wealth of evidence in the literature on Regge theory, for the existence of a
``soft pomeron" regime, where Regge behavior is determined by the dynamics of confinement.  Our methods, as yet,
give no clues about how to understand that regime.

\section{Acknowledgments} Special thanks go to Jo\~ao Penedones whose many comments and suggestions helped us sharpen our understanding of many issues. We also thank Juan Maldacena and Joe Polchinski for enlightening discussions and comments and John Mason for his input during the initial stage of this project. This work was supported in part by the U.S. Department of Energy grant number: DE-FG03-92ER40689.

\appendix

\section{An Integral}
The following integral can be found in the literature\cite{integrals,BFKL}:
\begin{eqnarray}
&&\int d^2 z^{a-1} (1-z)^{b-a-1}(1- z y)^{-c} \bar{z}^{\tilde{a}-1}(1-\bar{z})^{\tilde{b}-\tilde{a}-1}(1-\bar{z}\bar{y})^{-\tilde{c}}
\end{eqnarray}
For $\tilde{a}-a,\tilde{b}-b,\tilde{c}-c\in Z$ and $a,b,c,a-b,a-c,b-c\notin Z$ the result is:
\begin{eqnarray}
&&2i \lambda_0(a,b,\tilde{a},\tilde{b}) {_2}F_1(c,a;b;y){_2}F_1(\tilde{c},\tilde{a};\tilde{b};\bar{y})+2i \lambda_1(a,b,c,\tilde{a},\tilde{b},\tilde{c}) (y)^{1-b}(\bar{y})^{1-\tilde{b}}\cdot\\ &&\cdot {_2}F_1(c-b+1,a-b+1;2-b;y){_2}F_1(\tilde{c}-\tilde{b}+1,\tilde{a}-\tilde{b}+1;2-\tilde{b};\bar{y})
\nonumber
\end{eqnarray}
where
\begin{eqnarray}
&&\lambda_0(a,b,\tilde{a},\tilde{b})={\Gamma(a)\Gamma(b-a)\over \Gamma(b)}{\Gamma(\tilde{a})\Gamma(\tilde b-\tilde a)\over \Gamma(\tilde b)}{\sin \pi \tilde a \sin \pi(\tilde b-\tilde a)\over \sin \pi \tilde{b}}\\
&&\lambda_1(a,b,c,\tilde a, \tilde b,\tilde c)=-(-1)^{b-\tilde b +a-\tilde a}{\Gamma(b-1)\Gamma(1-c)\over \Gamma(b-c)}{\Gamma(\tilde b-1)\Gamma(1-\tilde c)\over \Gamma (\tilde b-\tilde c)}{\sin \pi b \sin \pi c\over \sin \pi(b-c)}\nonumber
\end{eqnarray}


\begin{thebibliography}{19}





\bibitem{sum}
  R.~Blankenbecler and R.~L.~Sugar,
  Phys.\ Rev.\  {\bf 168}, 1597 (1968).
 J.~L.~Cardy,
  Phys.\ Lett.\  B {\bf 53}, 355 (1974).
 C.~Lovelace,
  Phys.\ Lett.\  B {\bf 55}, 187 (1975).
 J.~B.~Bronzan and R.~L.~Sugar,
  Phys.\ Rev.\  D {\bf 17}, 585 (1978).
  C.~Gilain,
 M.~C.~Bergere and C.~Gilain,
  J.\ Math.\ Phys.\  {\bf 19}, 1495 (1978).
 J.~B.~Bronzan and R.~L.~Sugar,
  Phys.\ Rev.\  D {\bf 17}, 2813 (1978).
 M.~C.~Bergere and C.~de Calan,
  Phys.\ Rev.\  D {\bf 20}, 2047 (1979).
 S.~J.~Chang and T.~M.~Yan,
  Phys.\ Rev.\  D {\bf 7}, 3698 (1973).
 J.~L.~Gervais and A.~Neveu,
  Nucl.\ Phys.\  B {\bf 163}, 189 (1980).
 H.~Cheng and T.~T.~Wu,
  Phys.\ Rev.\  D {\bf 5}, 3189 (1972).
 H.~Cheng and T.~T.~Wu,
  Phys.\ Rev.\  D {\bf 5}, 3192 (1972).
 S.~G.~Matinyan and A.~G.~Sedrakian,
  Yad.\ Fiz.\  {\bf 24}, 844 (1976).
  I.~I.~Balitsky, L.~N.~Lipatov and V.~S.~Fadin,
{\it  In *Leningrad 1979, Proceedings, Physics Of Elementary Particles*, Leningrad 1979, 109-149}
 D.~A.~Dicus, D.~Z.~Freedman and V.~L.~Teplitz,
  Phys.\ Rev.\  D {\bf 4}, 2320 (1971).
 T.~Jaroszewicz,
  Acta Phys.\ Polon.\  B {\bf 11}, 965 (1980).
 M.~Praszalowicz,
  Acta Phys.\ Polon.\  B {\bf 12}, 773 (1981).
 H.~M.~Fried,
  Phys.\ Rev.\  D {\bf 1}, 596 (1970).
 D.~Shapero,
  Phys.\ Rev.\  {\bf 186}, 1697 (1969).
 B.~W.~Lee and R.~F.~Sawyer,
  Phys.\ Rev.\  {\bf 127}, 2266 (1962).
 R.~A.~Brandt and C.~A.~Orzalesi,
  Phys.\ Lett.\  B {\bf 34}, 641 (1971).
 H.~B.~Nielsen,
  Phys.\ Lett.\  B {\bf 35}, 515 (1971).
 G.~V.~Frolov, V.~N.~Gribov and L.~N.~Lipatov,
  Phys.\ Lett.\  B {\bf 31}, 34 (1970).
 K.~Seto,
  Prog.\ Theor.\ Phys.\  {\bf 40}, 605 (1968).

\bibitem{BFKL}
  E.~A.~Kuraev, L.~N.~Lipatov and V.~S.~Fadin,
  Sov.\ Phys.\ JETP {\bf 44}, 443 (1976)
  [Zh.\ Eksp.\ Teor.\ Fiz.\  {\bf 71}, 840 (1976)].

   I.~I.~Balitsky and L.~N.~Lipatov,
  Sov.\ J.\ Nucl.\ Phys.\  {\bf 28}, 822 (1978)
  [Yad.\ Fiz.\  {\bf 28}, 1597 (1978)].

  L.~N.~Lipatov,
  Sov.\ Phys.\ JETP {\bf 63}, 904 (1986)
  [Zh.\ Eksp.\ Teor.\ Fiz.\  {\bf 90}, 1536 (1986)].

\bibitem{mueller}
  A.~H.~Mueller,
  Nucl.\ Phys.\  B {\bf 415}, 373 (1994).

   A.~H.~Mueller,
  Nucl.\ Phys.\  B {\bf 437}, 107 (1995)
  [arXiv:hep-ph/9408245].

\bibitem{bpst}
  R.~C.~Brower, J.~Polchinski, M.~J.~Strassler and C.~I.~Tan,
  JHEP {\bf 0712}, 005 (2007)
  [arXiv:hep-th/0603115].

\bibitem{Maldacena:1997re}
  J.~M.~Maldacena,
  Adv.\ Theor.\ Math.\ Phys.\  {\bf 2}, 231 (1998)
  [Int.\ J.\ Theor.\ Phys.\  {\bf 38}, 1113 (1999)]
  [arXiv:hep-th/9711200].

  S.~S.~Gubser, I.~R.~Klebanov and A.~M.~Polyakov,
  Phys.\ Lett.\  B {\bf 428}, 105 (1998)
  [arXiv:hep-th/9802109].

\bibitem{port}
  L.~Cornalba, M.~S.~Costa and J.~Penedones,
  JHEP {\bf 0806}, 048 (2008)
  [arXiv:0801.3002 [hep-th]].

  L.~Cornalba, M.~S.~Costa, J.~Penedones and R.~Schiappa,
  JHEP {\bf 0708}, 019 (2007)
  [arXiv:hep-th/0611122].

  L.~Cornalba, M.~S.~Costa, J.~Penedones and R.~Schiappa,
  Nucl.\ Phys.\  B {\bf 767}, 327 (2007)
  [arXiv:hep-th/0611123].

  L.~Cornalba, M.~S.~Costa and J.~Penedones,
  arXiv:0911.0043 [hep-th].


\bibitem{mh}
  D.~M.~Hofman and J.~Maldacena,
  JHEP {\bf 0805}, 012 (2008)
  [arXiv:0803.1467 [hep-th]].

\bibitem{do}
  F.~A.~Dolan and H.~Osborn,
  Nucl.\ Phys.\  B {\bf 599}, 459 (2001)
  [arXiv:hep-th/0011040].

  F.~A.~Dolan and H.~Osborn,
  Nucl.\ Phys.\  B {\bf 678}, 491 (2004)
  [arXiv:hep-th/0309180].

\bibitem{ln}
  F.~E.~Low,
  Phys.\ Rev.\  D {\bf 12}, 163 (1975).

  S.~Nussinov,
  Phys.\ Rev.\  Lett. {\bf 34}, 1286 (1975).

\bibitem{symbfkl}
  A.~V.~Kotikov and L.~N.~Lipatov,
  Nucl.\ Phys.\  B {\bf 661}, 19 (2003)
  [Erratum-ibid.\  B {\bf 685}, 405 (2004)]
  [arXiv:hep-ph/0208220].

  A.~V.~Kotikov and L.~N.~Lipatov,
  Nucl.\ Phys.\  B {\bf 582}, 19 (2000)
  [arXiv:hep-ph/0004008].

\bibitem{balitsky} I.~Balitsky,
  arXiv:hep-ph/0101042.

\bibitem{integrals}
  J.~S.~Geronimo and H.~Navelet,
  J.\ Math.\ Phys.\  {\bf 44}, 2293 (2003)
  [arXiv:math-ph/0003019].

\end{thebibliography}
\end{document}